\documentstyle[aps,epsf,prl]{revtex}
\begin{document}

\title{Two-proton correlations from 158 AGeV Pb+Pb central collisions}

\author{
H.~Appelsh\"{a}user$^{7,\#}$, J.~B\"{a}chler$^{5}$,
S.J.~Bailey$^{17}$, D.~Barna$^{4}$, L.S.~Barnby$^{3}$,
J.~Bartke$^{6}$, R.A.~Barton$^{3}$, L.~Betev$^{12}$, 
H.~Bia{\l}\-kowska$^{15}$, A.~Billmeier$^{10}$
% ...
}
\author{(The NA49 collaboration)}

\address{
% ...
$^{3}$Birmingham University, Birmingham, England.\\
$^{4}$KFKI Research Institute for Particle and Nuclear Physics, 
	Budapest, Hungary.\\
$^{5}$CERN, Geneva, Switzerland.\\
$^{6}$Institute of Nuclear Physics, Cracow, Poland.\\
$^{7}$Gesellschaft f\"{u}r Schwerionenforschung (GSI), Darmstadt, Germany.\\
% ...
$^{10}$Fachbereich Physik der Universit\"{a}t, Frankfurt, Germany.\\
% ...
$^{12}$University of California at Los Angeles, Los Angeles, CA, USA.\\
% ...
$^{15}$Institute for Nuclear Studies, Warsaw, Poland.\\
% ...
$^{17}$Nuclear Physics Laboratory, University of Washington, 
	Seattle, WA, USA.\\
% ...
$^{\#}$present address: Physikalisches Institut, Universitaet Heidelberg, 
	Germany.\\
% ...
}

\maketitle

\begin{abstract}
The two-proton correlation function at midrapidity from Pb+Pb central
collisions at 158 AGeV has been measured by the NA49 experiment.
The results are compared to model predictions from static thermal
Gaussian proton source distributions and transport models {\sc rqmd}
and {\sc venus}. An effective proton source size is determined by
minimizing $\chi^2/{\rm ndf}$ between the correlation functions of the
data and those calculated for the Gaussian sources, yielding
$\sigma_{\rm eff}=3.85\pm 0.15{\rm (stat.)} ^{+0.60}_{-0.25}{\rm (syst.)}$~fm.
Both the {\sc rqmd} and the {\sc venus} model are consistent 
with the data within the error in the correlation peak region.
\end{abstract}

\twocolumn

Nuclear matter at high energy density has been extensively studied
through high energy heavy ion collisions (for recent developments,
see~\cite{qm}). It is hoped that these collisions will create a
deconfined state, the quark-gluon plasma.
These studies have shown that nucleus-nucleus collisions are not
mere superpositions of nucleon-nucleon collisions.
In Pb+Pb central collisions at the CERN SPS, the nucleons are transported 
further into the midrapidity region than in S+S central collisions and 
p+p interactions
\cite{App98:na49_prl_baryon,Bea96:na44_plb,Alb98:na35_epj,Agu91:pp},
consistent with the picture that the incoming nucleons in nucleus-nucleus
collisions have undergone a significant number of scatterings.
This results in a high rapidity density of net-baryons at
midrapidity~\cite{App98:na49_prl_baryon}.
The spatial baryon density plays an important role in the 
dynamical evolution of heavy ion collisions~\cite{Koc83Lee88}.
An essential ingredient for establishing the spatial baryon density
is the space-time extent of the baryon source at freezeout, 
which can be inferred from two-proton correlation
functions~\cite{Koo77:plb_pp,Led82Gel90}.

Two-proton correlations at low relative momentum are due to final state 
interactions (the attractive strong and the repulsive Coulomb interactions)
and Fermi-Dirac quantum statistics~\cite{Koo77:plb_pp,Led82Gel90}.
The correlation function is zero at 
$q_{\rm inv}=\sqrt{-q_{\mu}q^{\mu}}/2 = 0$ and peaks at 
$q_{\rm inv}\approx 20$~MeV/$c$, where $q_{\mu}$ is the difference
of the proton 4-momenta, and $q_{\rm inv}$ is the momentum 
magnitude of one proton in the rest frame of the pair.
The peak height is inversely related to the space-time extent
of the proton source~\cite{Koo77:plb_pp,Led82Gel90}.

The space-time extent of the pion source has been studied extensively
via two-pion interferometry in heavy ion collisions 
\cite{App98:epj_expansion,hbt_ref1,hbt_ref2}.
However, measurements of the space-time extent of the proton source
are rare, especially for high energy nucleus-nucleus 
collisions~\cite{two_proton}.
In this letter, we report the first measurement of the two-proton 
correlation function in the midrapidity region from Pb+Pb central 
collisions at 158 AGeV, by the NA49 collaboration at the CERN SPS.

NA49 is a large acceptance hadron spectrometer~\cite{na49_nim}.
The main detectors are four large time-projection chambers (TPCs). 
Two TPCs (VTPCs) are placed along the beam axis inside two dipole magnets, 
which have a maximum integrated field strength of 9 Tesla-meters.
The other TPCs (MTPCs) are placed downstream of the magnets on 
either side of the beam axis.
Behind the MTPCs are two time-of-flight (TOF) walls, 
covering smaller acceptance than the TPCs.
A beam of $^{208}$Pb struck a Pb target of thickness 224 mg/cm$^2$,
placed in front of the first VTPC.
A zero-degree calorimeter, located further downstream on the axis of the 
deflected beam, measures the kinetic energy of the projectile spectators.
By requiring less than 6 TeV energy measured in the calorimeter, 
the 5\% most central events were selected.
These events correspond approximately to collisions with impact 
parameter $b\; ^{<}_{\sim}\; 3.3$~fm.

Two independent analyses of the two-proton correlation function 
were carried out:
\begin{enumerate}
\item $dE/dx$ analysis: 
Track segments reconstructed in the MTPCs were matched to 
track segments from at least one of the VTPCs.
Particle momenta were determined by track curvature in the VTPCs.
The particle identification was performed by measuring the specific 
ionization ($dE/dx$) deposited by a charged particle in the gas of 
the MTPC, measured in the region of the ``relativistic rise''.
The mean $\langle dE/dx \rangle$ of a particle was estimated 
using the truncated mean technique~\cite{trunc_mean}.
A relative $\langle dE/dx \rangle$ resolution of $\sigma\simeq 5$\%
was achieved in this analysis. 
Since the proton $\langle dE/dx \rangle$ is approximately $1\sigma$ 
and $3\sigma$ below those of the kaons and the pions with
the same momentum, protons cannot be uniquely identified.
Instead, particles with at least 70\% probability 
to be a proton were included in the analysis.
\item TOF analysis: 
Only tracks reconstructed in the MTPCs were used. 
The momentum of a particle was determined by an iterative procedure 
in which the trajectory fitted to the measured points was projected 
upstream through the magnetic field, assuming that it originated 
at the primary interaction vertex.
The particle identification was performed by combining the momentum 
measurement with the velocity information from the TOF walls and 
the $\langle dE/dx \rangle$ from the MTPCs. 
The typical TOF resolution was measured to be 60 ps.
\end{enumerate}

\begin{figure}[hbt]
\centerline{\epsfxsize=4.5in\epsfbox[0 260 560 600]{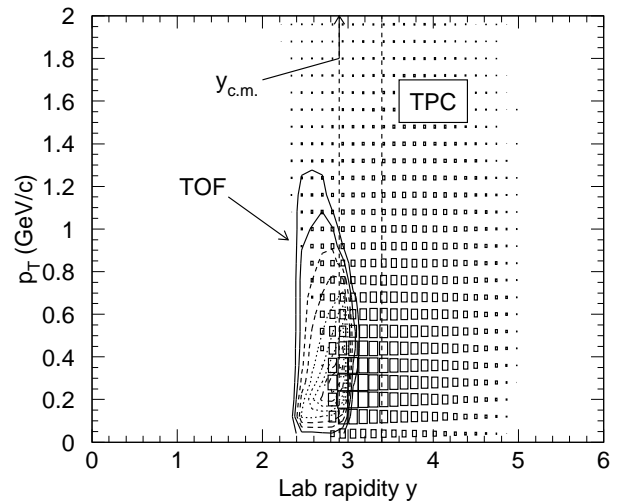}}
\caption{NA49 proton acceptance in transverse momentum $p_T$ versus 
	rapidity $y$. The acceptance values for protons used in 
	the $dE/dx$ analysis are shown in the box plot between the 
	dashed lines, and for the protons used in the TOF analysis 
	are shown in the contours.}
\label{fig:acc}
\end{figure}

Fig.~\ref{fig:acc} shows the proton acceptances from the two analyses,
which are in different hemispheres with respect to midrapidity 
($y_{\rm c.m.}=2.9$).
The two-proton correlation function was analysed using protons 
from the rapidity range $2.9<y<3.4$ in the $dE/dx$ analysis, 
and $2.4<y<2.9$ in the TOF analysis.
Both analyses used protons up to a transverse momentum $p_T=2$ GeV/$c$.

The two-proton correlation function was calculated as the ratio of 
the $q_{\rm inv}$ distribution of true proton pairs to that
of mixed-event pairs with protons taken from different events.
The number of mixed-event pairs is large enough that
the statistical error on the correlation function is dominated
by the statistical uncertainty in the number of true pairs.
To eliminate the effect of close pair reconstruction inefficiency,
a cut of 2~cm was applied on the pair distance at the mid-plane of
the MTPC for both true and mixed-event pairs~\cite{theses}.

The proton sample is contaminated by protons from weak decays
($\Lambda$ including feeddowns, and $\Sigma^+$) which are incorrectly 
reconstructed as primary vertex tracks. 
From the measured single particle 
distributions~\cite{App98:na49_prl_baryon,Bor97:sqm97} as well as
model calculations of {\sc rqmd} and {\sc venus}, we estimate 
the contamination to be $25 \pm 10\%$ (with weak $p_T$ dependence).
This results in $44^{+14}_{-16}\%$ of the proton pairs having at least 
one proton from weak decays, which is assumed to have no correlation 
with the protons produced in the primary interaction~\cite{markus_thesis}.
Hence, protons from weak decays only reduce the correlation strength,
without changing the shape of the two-proton correlation function at
low $q_{\rm inv}$.
Additional contamination is present in the $dE/dx$ analysis from kaons on 
the lower tail of their $\langle dE/dx \rangle$ distribution merging with 
the region where particles have at least 70\% probability to be a proton.
This results in 25\% $K^+p$ pairs and fewer than 2\% $K^+K^+$ pairs
in the proton pair sample, independent of $q_{\rm inv}$.
$K^+p$ pairs with $0<k<100$~MeV/$c$ 
($k$ is the momentum of the $K^+$ or proton in the rest frame of the pair),
which spans the region affected by Coulomb repulsion for the typical
source created at midrapidity in Pb+Pb central collisions,
affect the two-proton correlation function in the 
range $160<q_{\rm inv}<460$~MeV/$c$.
Only $K^+p$ pairs with $180<k<240$~MeV/$c$ contribute to the
two-proton correlation function in the range $0<q_{\rm inv}<50$~MeV/$c$.
Therefore, the $K^+p$ contamination in the proton pair sample shows
no structure in $q_{\rm inv}$ below 160~MeV/$c$, but reduces the 
correlation strength by 25\%.
The contribution from the $K^+K^+$ pairs in the two-proton correlation 
function is negligible.

\begin{figure}[hbt]
\centerline{\epsfxsize=4.5in\epsfbox[0 140 560 640]{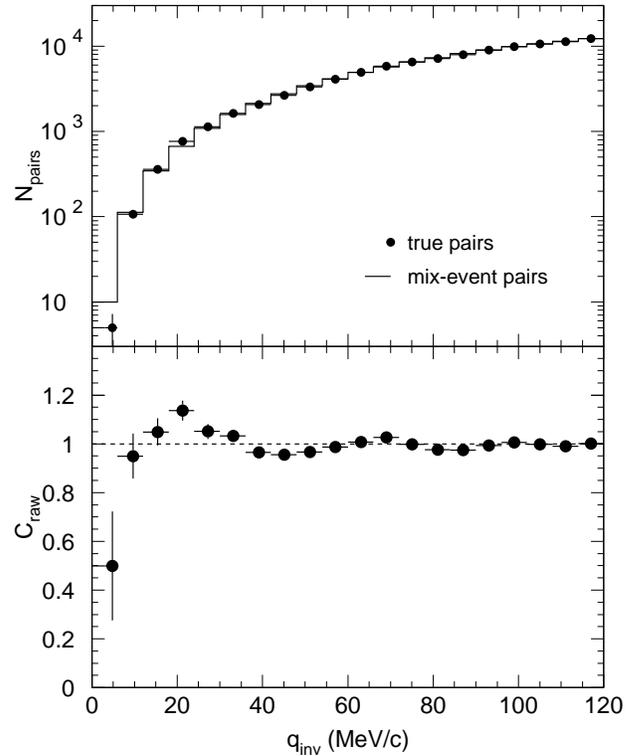}}
\caption{Upper panel: 
	$q_{\rm inv}$ distributions of true proton pairs (points) 
	and mixed-event proton pairs (histogram).
	Lower panel: 
	the measured two-proton correlation function.
	In both panels, the points are plotted at the mean 
	$q_{\rm inv}$ of each bin. The bin size is 6~MeV/$c$.
	The errors shown are statistical only.
	The contamination from weak decay protons and the finite
	$q_{\rm inv}$ resolution are not corrected.}
\label{fig:data}
\end{figure}

The correlation functions obtained from the two analyses can be 
directly compared because of the nearly symmetric acceptances used.
The corrected correlation functions are consistent.
In the results reported below, the corrected true pairs and 
the respective mixed-event pairs from the two analyses are combined.
The combined sample has about $10^5$ pairs with $q_{\rm inv}<120$ MeV/$c$,
75\% of which are from the TOF analysis.
The $q_{\rm inv}$ distributions of the true and the mixed-event pairs
are shown in the top panel of Fig.~\ref{fig:data}.
The number of mixed-event pairs is normalised to that of true pairs 
in the range $q_{\rm inv}>500$~MeV/$c$.
The resulting correlation function $C_{\rm raw}$ is shown in the 
lower panel of Fig.~\ref{fig:data}, and tabulated in Table~\ref{tab}.
The peak at $q_{\rm inv}\approx 20$~MeV/$c$ has amplitude $1.14\pm 0.04$.
A statistically significant structure is seen in the correlation
function at $q_{\rm inv}\approx 70$~MeV/$c$. 
The structure is present in both the $dE/dx$ and TOF data sets. 
See below for further discussion of the structure.

\begin{figure}[hbt]
\centerline{\epsfxsize=4in\epsfbox[0 190 560 590]{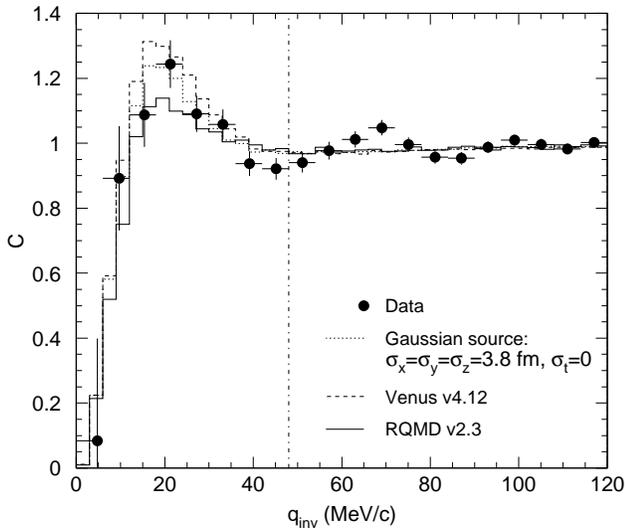}}
\caption{The two-proton correlation function after corrections for
	the 44\% contamination due to weak decay protons and the finite 
	$q_{\rm inv}$ resolution (points), compared to calculations for
	a Gaussian source (dotted histogram), and for freeze-out protons
	from {\sc rqmd} v2.3 (solid histogram) and {\sc venus} v4.12
	(dashed histogram). 
	The errors shown on the data points are statistical only.
	The dot-dashed line indicates the $q_{\rm inv}$ value (48~MeV/$c$)
	below which the comparisons between the data and the calculations
	are performed.}
\label{fig:model}
\end{figure}

The correlation function was further corrected for the contamination 
from weak decay protons and the finite $q_{\rm inv}$ resolution.
The former is corrected via $C^{\prime}=(C_{\rm raw}-0.44)/(1-0.44)$.
The momentum resolution is 0.25\% at a momentum about 10~GeV/$c$;
the $q_{\rm inv}$ resolution is 2.5~MeV/$c$ in the range 
$q_{\rm inv}<120$~MeV/$c$~\cite{theses}.
The effect of the $q_{\rm inv}$ resolution 
was studied using the {\sc rqmd} and {\sc venus} models, 
and was found to be significant only for the first 
two data points of the measured correlation function. 
The final corrected correlation function is plotted in 
Fig.~\ref{fig:model} as filled points, and tabulated in Table~\ref{tab}.

In order to assess the proton freeze-out conditions, we compare the 
measured two-proton correlation function to theoretical calculations.
Given the proton phase space density distribution, the two-proton 
correlation function can be calculated by the Koonin-Pratt 
Formalism~\cite{Koo77:plb_pp,Pra87:two_proton}.
The formalism uses the phase-shift method, which incorporates 
the Coulomb and the strong interactions between the protons.
We use two types of proton freeze-out distributions:

(I) Gaussian sources of widths $\sigma_{x,y}, \sigma_z$ and $\sigma_t$
for the space and time coordinates of protons in the source rest frame,
and a Boltzmann distribution with temperature $T$ for the proton 
momentum distribution.
No correlation between space-time and momentum is present.
The following combinations of parameters are used:
 $\sigma_{x,y}=\sigma_z=\sigma$; $\sigma_t=0$ and $\sigma$;
$T=120$~MeV (as derived in~\cite{App98:epj_expansion}), 
300~MeV (measured inverse slope of proton transverse mass 
spectrum~\cite{App98:na49_prl_baryon})
and 70~MeV (inverse slope observed at low energy, as an extreme).

(II) Protons generated for Pb+Pb central collisions ($b\leq 3.3$~fm)
at 158 AGeV by two microscopic transport models:
the {\sc rqmd} model (version 2.3)~\cite{rqmd} 
and the {\sc venus} model (version 4.12)~\cite{Wer93:venus}.
Both models describe a variety of experimental data 
on single particle distributions reasonably well.
Protons at freeze-out have correlations between space-time
and momentum intrinsic to the dynamical evolution in the models.
Neither model includes correlations due to quantum statistics and
final state interactions.
Particle weak decays are not included in the models.

Only protons in the experimental acceptance are used
to calculate the two-proton correlation functions,
the results of which are shown in Fig.~\ref{fig:model}. 
The correlation function for the Gaussian source 
with $\sigma_{x,y}=\sigma_z=3.8$~fm, $\sigma_t=0$~fm and $T=120$~MeV,
which describes well the experimental data, is also plotted.

We use $\chi^2$/ndf, the normalised mean square of the point-to-point
difference between the data and the calculation in the range 
$q_{\rm inv}<48$~MeV/$c$ ({\it i.e.}, 8 data points or ${\rm ndf}=8$),
to quantify how well the calculations agree with the data.
We characterize the effective size of the model-generated proton source by
 $\sigma_{\rm eff} = 
 \sqrt[3]{\sigma_{\Delta x}\cdot\sigma_{\Delta y}\cdot\sigma_{\Delta z}}/
 \sqrt{2}$,
where $\sigma_{\Delta x}, \sigma_{\Delta y}$ and $\sigma_{\Delta z}$ 
are the Gaussian widths fitted to the distributions 
in $\Delta x, \Delta y$ and $\Delta z$, the distance between 
the protons of close pairs with $q_{\rm inv}<48$~MeV/$c$.
The distance is evaluated in the pair rest frame (since the correlation
function is studied as a function of $q_{\rm inv}$) at the time 
when the later particle freezes out~\cite{note_2m}.
Repectively for {\sc rqmd} and {\sc venus}, the $\chi^2$/ndf values 
are 1.53 and 1.38, corresponding to a probability of 20\% and 14\% 
that the model distributions are consistent with the data; 
the fitted Gaussian widths are
($\sigma_{\Delta x},\sigma_{\Delta y},\sigma_{\Delta z})=(5.91,6.00,6.83)$~fm
and $(4.57,4.57,6.08)$~fm, where $z$ is the longitudinal coordinate;
consequently the effective sizes are $\sigma_{\rm eff}=4.41$~fm 
and $3.55$~fm.

\begin{figure}[hbt]
\centerline{\epsfxsize=4in\epsfbox[0 190 560 600]{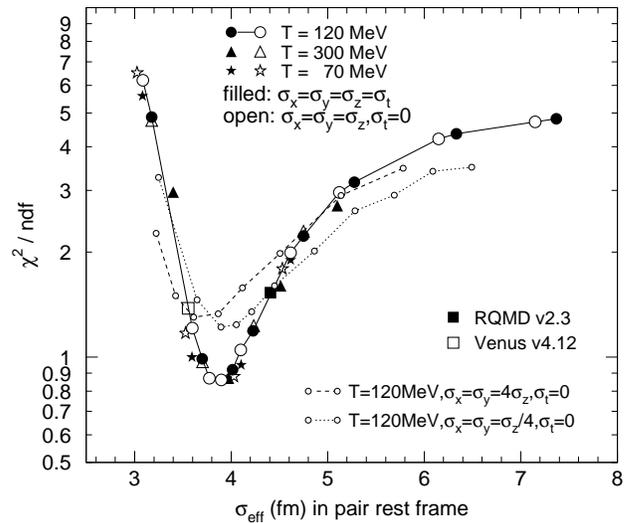}}
\caption{The $\chi^2$/ndf values as function of the effective source
	size $\sigma_{\rm eff}$ for various model calculations with 
	respect to the measured correlation function. 
	The model calculations are for Gaussian sources 
	(circles, triangles, and stars, both filled-in and open) 
	and for freeze-out protons generated by {\sc rqmd} v2.3 
	(filled square) and {\sc venus} v4.12 (open square). 
	The curves connect the various sets of points to guide the eye.
	The errors on $\sigma_{\rm eff}$ are negligible for all the points.}
\label{fig:chi2}
\end{figure}

In Fig.~\ref{fig:chi2}, we study the $\chi^2$/ndf as a function
of $\sigma_{\rm eff}$.
The $\chi^2$/ndf values for all three models follow roughly the same
solid curve, drawn through the points for the Gaussian sources 
with $T=120$~MeV to guide the eye~\cite{note_curve}.
From the minimum $\chi^2$/ndf (=0.86) point and the points 
where $\chi^2$/ndf has increased by 0.125 (note ${\rm ndf}=8$), 
we extract $\sigma_{\rm eff}=(3.85\pm 0.15)$~fm,
where 0.15~fm is the statistical error~\cite{PDG98:error}.
By applying a correction to the measured correlation function 
using a proton pair contamination of 28\% and 58\%, we obtain 
a systematic error of $\pm 0.15$~fm on $\sigma_{\rm eff}$.
We note that $\sigma_{\rm eff}=3.85$~fm corresponds to a uniform 
density hard sphere of radius $\sqrt{5}\sigma_{\rm eff}=8.6$~fm, 
which is larger than the size of the colliding Pb nuclei.

We have also studied Gaussian sources with extreme shapes: 
oblate $\sigma_{x,y}=4\sigma_z$ and 
prolate $\sigma_{x,y}=\sigma_z/4$ (both with $\sigma_t=0$).
The corresponding $\chi^2$/ndf versus $\sigma_{\rm eff}$ curves
are shown in Fig.~\ref{fig:chi2} as dashed and dotted lines, respectively. 
They do not fall along the curve for the isotropic Gaussian sources,
implying that the two-proton correlation function has certain 
sensitivity to the shape of the proton source.
The data do not favour the extreme Gaussian shapes.
In principle, multi-dimensional two-proton correlation functions
could reveal the shape of the proton 
source~\cite{Koo77:plb_pp,Pra87:two_proton}. 
However, this requires greater statistical precision 
than is available with the present data set.

By minimizing the $\chi^2$/ndf between the correlation functions for 
the various Gaussian sources with respect to those for {\sc rqmd} and 
{\sc venus}, we can extract the effective proton source size for the models. 
The results are consistent with the calculated effective sizes 
from the models themselves.

We note that there is no simple relation between $\sigma_{\rm eff}$
and the proton source in the models.
From single proton distributions at freeze-out, we obtain 
the following Gaussian widths in the source rest frame:
 ($\sigma_x,\sigma_y,\sigma_z,\sigma_t)=(7.6,7.7,6.4,7.0)$~fm
for {\sc rqmd} and $(3.6,3.6,4.3,1.9)$~fm for {\sc venus}, respectively.
The two-proton correlation function, therefore, appears to measure 
a smaller region of the source due to space-time-momentum correlation.
The effect is more dramatic in {\sc rqmd} than in {\sc venus}, 
which is consistent with the expectation that more secondary 
particle interactions in {\sc rqmd} result in a stronger correlation 
between space-time and momentum of freeze-out protons.
The fact that the $\chi^2$/ndf values versus $\sigma_{\rm eff}$ 
for the models lie on the curve obtained for the Gaussian sources,
in which no space-time-momentum correlation is present,
suggests that the effect of the space-time-momentum correlation 
is small in $\sigma_{\rm eff}$. 

Now we come back to the structure observed at 
$q_{\rm inv}\approx 70$~MeV/$c$ in the correlation function.
Many systematics have been checked, including
(1) $p_T$ dependence,
(2) stricter acceptance cut,
(3) larger cut on two track distance,
(4) varying cuts to allow different kaon contaminations 
	in the TOF data set~\cite{markus_thesis},
(5) allowing only one particle from the same event in event mixing,
(6) residual effect in single particle spectra 
	due to two-particle correlation,
(7) misreconstructed momenta for $\Lambda$ decay protons, and 
(8) reflection of $\Lambda p$ strong interactions in two-proton
correlation with one proton from $\Lambda$ decays~\cite{Wan99}.
None of the checked systematics can account for the structure.
Nevertheless, if we assume that the structure has an oscillatory
behavior, $\cos(aq_{\rm inv})/q_{\rm inv}$ (where $a$ is a constant), 
and also contributes to the correlation peak region, 
then we estimate a systematic error of $^{+0.45}_{-0.10}$~fm 
in the extracted $\sigma_{\rm eff}$, 
in addition to the $\pm 0.15$~fm estimated above.

It is, however, possible that the structure is due to underlying physics.
For instance, such structure may result from a breakdown of the 
basic assumptions underlying the Koonin-Pratt Formalism~\cite{Bro99}: 
the smoothness assumption~\cite{Pra97}, 
the independent emission assumption, and/or
the $q_{\rm inv}$ independence of the two-proton source. 
A hard edge in the two-proton source distribution may also result in
similar structures, however, it does not reproduce the observed sign 
and amplitude, unless the independent emission assumption is violated.
See also the discussion in Ref.~\cite{Bro99}.

Finally, we comment on our two-proton correlation function
in the context of other measurements.
The pion source size measured by interferometry increases 
with the pion multiplicity~\cite{hbt_ref2},
which increases steadily with bombarding energy 
in similar colliding systems~\cite{Gaz95:pion}.
Due to the large pion-nucleon cross-section, one would expect that 
protons and pions freeze-out under similar conditions, therefore, 
the proton source size would increase with bombarding energy as well.
However, our measurement, in conjunction with preliminary results 
obtained at GSI~\cite{gsi} and AGS~\cite{e895} energies, shows 
that the peak height is rather insensitive to the bombarding energy.
This implies that the effective sizes of the freeze-out proton sources 
are similar in heavy ion collisions over a wide energy range.

In summary, NA49 has measured the two-proton correlation function at 
midrapidity from Pb+Pb central collisions at 158 AGeV.
The peak height of the two-proton correlation function 
is $1.24 \pm 0.07$ after corrections.
From comparison between the data and the calculations for static 
thermal Gaussian sources, we extract an effective proton source size
$\sigma_{\rm eff}=3.85\pm 0.15{\rm (stat.)} ^{+0.60}_{-0.25}{\rm (syst.)}$~fm.
Within the error, the {\sc rqmd} model ($\sigma_{\rm eff}=4.41$~fm) and the 
{\sc venus} model ($\sigma_{\rm eff}=3.55$~fm) are consistent with the data.
Due to the space-time-momentum correlation, the two-proton correlation 
function is sensitive only to a limited region of the proton source.
Our measurement together with the measurements at lower energies suggest
a very weak dependence of the two-proton correlation function on
bombarding energy. 
The observed structure at $q_{\rm inv}\approx 70$~MeV/$c$ is not understood.

We thank Dr. H. Sorge, Dr. K. Werner, and Dr. S. Pratt 
for providing us the theoretical models.
This work was supported by the Director, Office of Energy Research, 
Division of Nuclear Physics of the Office of High Energy and Nuclear 
Physics of the US Department of Energy under Contract DE-AC03-76SF00098, 
the US National Science Foundation, 
the Bundesministerium fur Bildung und Forschung, Germany, 
the Alexander von Humboldt Foundation, 
the UK Engineering and Physical Sciences Research Council, 
the Polish State Committee for Scientific Research (2 P03B 01912 and 9913), 
the Hungarian Scientific Research Foundation 
under contracts T14920 and T23790,
the EC Marie Curie Foundation, the Polish-German Foundation,
and the Czech Republic under Contract 202/98/1283.

\begin{table}
\caption{The two-proton correlation function.
	$\langle q_{\rm inv}\rangle$: mean $q_{\rm inv}$ in MeV/$c$;
	$C_{\rm raw}$: uncorrected correlation function;
	$C$: correlation function corrected for 44\%
	contamination in the proton pair sample due to weak 
	decay protons and for the finite $q_{\rm inv}$ resolution.
	All quoted errors are statistical only.}
\label{tab}
\begin{tabular}{r|c|c}
	$\langle q_{\rm inv}\rangle$ & $C_{\rm raw}$ & $C$ \\ \hline
   4.8 & $0.500 \pm 0.224$ & $0.084 \pm 0.314$ \\
   9.7 & $0.949 \pm 0.092$ & $0.892 \pm 0.161$ \\
  15.4 & $1.049 \pm 0.055$ & $1.087 \pm 0.099$ \\
  21.3 & $1.136 \pm 0.041$ & $1.244 \pm 0.074$ \\
  27.2 & $1.051 \pm 0.031$ & $1.090 \pm 0.056$ \\
  33.2 & $1.033 \pm 0.026$ & $1.058 \pm 0.046$ \\
  39.1 & $0.965 \pm 0.021$ & $0.937 \pm 0.038$ \\
  45.1 & $0.956 \pm 0.019$ & $0.921 \pm 0.033$ \\
  51.1 & $0.966 \pm 0.017$ & $0.940 \pm 0.030$ \\
  57.1 & $0.987 \pm 0.015$ & $0.977 \pm 0.028$ \\
  63.1 & $1.007 \pm 0.014$ & $1.012 \pm 0.026$ \\
  69.1 & $1.027 \pm 0.013$ & $1.048 \pm 0.024$ \\
  75.1 & $0.998 \pm 0.012$ & $0.996 \pm 0.022$ \\
  81.1 & $0.976 \pm 0.012$ & $0.958 \pm 0.021$ \\
  87.1 & $0.974 \pm 0.011$ & $0.954 \pm 0.019$ \\
  93.0 & $0.993 \pm 0.010$ & $0.988 \pm 0.019$ \\
  99.0 & $1.005 \pm 0.010$ & $1.010 \pm 0.018$ \\
 105.0 & $0.998 \pm 0.010$ & $0.996 \pm 0.017$ \\
 111.0 & $0.990 \pm 0.009$ & $0.982 \pm 0.017$ \\
 117.0 & $1.001 \pm 0.009$ & $1.002 \pm 0.016$
\end{tabular}
\end{table}


\begin{thebibliography}{10}
\bibitem{note_2m}
For a chaotic Gaussian source, $\sigma_{\rm eff}$ is close to the
non-relativistic approximation of the effective size in the source 
rest frame, $\sqrt[6]{(\sigma_{x,y}^2+\frac{T}{2m}\sigma_t^2)^2\cdot
(\sigma_z^2+\frac{T}{2m}\sigma_t^2)}$, where $\frac{T}{m}$ is the
one-dimensional mean squared velocity of the thermal protons, 
and $\frac{T}{2m}$ is that of the close pairs.
\bibitem{note_curve}
The shape of the $\chi^2/{\rm ndf}$ versus $\sigma_{\rm eff}$ curves can 
be understood as follows: the $\chi^2$/ndf approaches an asymptotic value
for large sources to which the two-proton correlation function is not 
sensitive to the source size, whereas for small sources, the
two-proton correlation function is very sensitive and its strength 
decreases rapidly with increasing source size.
\end{thebibliography}
\end{document}